\documentclass[12pt]{article}
\usepackage{graphicx}

\def\pt{$p_T$}
\def\dis{distribution}
\def\az{azimuthal}

\def\mm{momentum}

\begin{document} 
\begin{center}  {\Large {\bf Elliptic Flow arising from  Ridges\\ due to Semi-hard Scattering}}
\vskip .75cm
 {\bf Rudolph C. Hwa}
\vskip.5cm {Institute of Theoretical Science and Department of Physics\\ University of Oregon, Eugene, OR 97403-5203, USA}
\end{center}
\vskip.5cm

\begin{abstract}
Azimuthal anisotropy in heavy-ion collisions is studied by taking into account the ridges generated by semi-hard scattering of intermediate-momentum partons, which can be sensitive to the initial spatial configuration of the medium in non-central collisions. In a simple treatment of the problem where the recombination of only thermal partons is considered, analytical formulas can be derived that yield results in accord with the data on $v_2$ for $p_T <$1-2 GeV/c. 
Centrality dependence is described by a geometrical factor. Ridge phenomenology is used to determine the initial slopes of $v_2$ at low \pt\ for both pion and proton. For higher \pt, shower partons from high-\pt\ jets must be included, but they are not considered here. 
\end{abstract}

\section{Introduction}

Hard scattering of partons is known to produce low-\pt\ hadrons by fragmentation, but it occurs so rarely that its influence on the bulk medium in heavy-ion collisions is not important. Semi-hard scattering, however, occurs much more frequently. Although there is no reliable way to calculate weak jets, their effect on low-\pt\ phenomenology should not be ignored. Recent experiments at RHIC have shown that hadrons associated with triggers at intermediate \pt\ exhibit ridge structure at low \pt\ that protrudes above the background \cite{ja,jp,jb}. Such structure has also been seen in auto-correlation without triggers \cite{tt}. The question that we address here is what aspect of the properties of the hadrons produced at low \pt\ can be influenced by the ridges generated by the semi-hard scattering. Since the time scale involved in such scattering is short, the ridges that are produced near the surface are sensitive to the initial configuration of the medium. We show  in this paper that such semi-hard scattering processes can drive the elliptic flow of the medium. The mechanism represents an alternative to the conventional hydrodynamical description that relies on rapid thermalization. Our approach does not replace hydrodynamics in general, but releases it from the burden of assuming its validity at early time.

We restrict our attention in this paper to the low $p_T$ region with $p_T < 2$ GeV/c.  The hadronization process will be treated by recombination, as we have done in the region $p_T > 2$ GeV/c \cite{hy}.  In the higher $p_T$ region the shower partons from high-$p_T$ jets must be considered; their contribution to $v_2$ will be investigated in a future study.  
That region is not germane to the issue at hand of relating the ridges to elliptic flow at low $p_T$ where only thermal partons are relevant. The ridges  we consider are generated by semi-hard jets that are produced much more copiously than the high-momentum jets.  The formation of ridges with or without identifiable peaks has been studied before in the recombination model \cite{ch}.  We now apply similar consideration to the study of azimuthal anisotropy. The relevance of recombination has been demonstrated by the observation of quark number scaling of $v_2$ \cite{mo,ad}.

\section{Ridge Properties}
Let us adopt the very simple description of nuclear geometry with the assumption that the colliding nuclei have sharp boundaries at the spherical surface of radius $R_A$.  At impact parameter $b$ the almond-shaped overlap region in the transverse plane is bounded by two circular arcs whose angular ranges are 
\begin{eqnarray}
\left| \phi \right| < \Phi \quad {\rm and} \quad  \left|\pi - \phi \right|< \Phi ,
\label{1}
\end{eqnarray}
where $\phi$ is the azimuthal angle measured from the centers of the two nuclei, respectively, and
\begin{eqnarray}
\Phi  = \cos^{-1} (b/2R_A) .
\label{2}
\end{eqnarray}
That is the spatial configuration of the system  at the initial time immediately after collision. The \mm\ of a detected hadron will have an \az\ angle also denoted by $\phi$. The expansion of the system at later times may well be described by hydrodynamics, say, for $\tau>3$ fm/c, so by the time of hadronization the medium  has characteristics of flow. Our emphasis here is to replace early time hydrodynamics by ridge phenomenology.

The issue concerning the $\phi$ dependence of the detected hadrons is about the physics that is sensitive to the initial configuration of the system.     If one considers semi-hard scattering as the  virtuality is lowered, there comes a region when the process is soft 
 enough to have frequent occurrences but hard enough to create intermediate-$p_T$ jets at early times.  If those jets are in the 2-3 GeV/c range, the corresponding time scale is $\tau<0.1$ fm/c, short enough to be sensitive to the initial spatial configuration.

   In events triggered by intermediate-$p_T$ particles  it has been found that jet structure   consists of a prominent peak above a ridge in the $\Delta \eta$ distribution \cite{ja}.  Quantitative study of the relative strength of the peak (denoted as Jet, or $J$) and the ridge ($R$) shows that $J/R$ becomes small when the trigger momentum is low and the associated-particle momentum is even lower ($<2$ GeV/c) \cite{jp,jb}.  When the $R$ yield dominates over the $J$ yield, the underlying jet is hardly recognizable and has been referred to as the phantom jet \cite{rch}, since it is a Jet-less jet.  The ridge  has been described as the recombination product of enhanced thermal partons due to the energy lost by the semi-hard parton traversing the medium \cite{ch}.  The ridge being in the immediate vicinity of the jet direction  implies that the semi-hard scattering takes place near the surface of the medium and that the thermalization of the energy lost is restricted to the region near the jet trajectory.

Applying the notion of phantom jets to our present problem, we have a collection of those semi-hard jets on the surface of the almond-shaped region, if the virtuality of the jets is small enough so that many of them are produced in a nuclear collisions. 
At $\eta\sim 0$ the semi-hard scattering involves low-$x$ $(<0.03)$ partons, which are abundant.
The direction of a scattered  parton is random, but in any given patch of spatial region near the surface the average direction of all outward partons is normal to the surface, since that  is the only direction in the geometry of the problem.  
When integrated over the surface, there is a layer of emitters of semi-hard partons whose directions are prescribed by the geometry of the overlap at the initial time.
The ridge of enhanced thermal partons may develop subsequently, after allowing some time for local thermalization to take place near the surface.  But the average directions of the phantom jets are  specified by Eq.\ (\ref{1}), independent of the time it takes for hadrons to form.   Since those phantom jets are the products of semi-hard partons directed outwards, the recoil partons directed inward are absorbed by the medium and contribute to the thermalization of the bulk that has no preferred direction of expansion in the transverse plane.  In the present study we ignore any aspect of the longitudinal expansion and consider only hadrons at mid-rapidity; thus the property of the ridges in $\eta$ is not as important as their $\phi$ distribution.
In summary  the system under study can be described by two main types of thermal partons that are generated by the collision:  (a) the bulk that is isotropic in $\phi$, and (b) the ridges  that is nonzero in the $\phi$ region  specified by Eq.\ (\ref{1}).

In Ref. \cite{hy} it is shown how the pion distribution can be obtained from the parton  distribution in the recombination model.  The region of interest there is for $p_T > 2$ GeV/c, so shower partons are important in addition to the thermal partons.  We now use the same formalism for $p_T < 2$ GeV/c and consider only the thermal partons.  For the bulk medium the invariant distribution of the thermal partons just before hadronization is
\begin{eqnarray}
q_0 {dN_q^B \over dq_Td\phi} = Cq_Te^{-q_T/T}
\label{3}
\end{eqnarray}
where, with our attention focused on only the momentum in the transverse plane at $y = 0$, $q_T$ is the parton transverse momentum and $q_0$ is its energy.  The value of $C$  is of no concern here since it will be canceled in a ratio to be calculated. 
  Adopting the same formalism of recombination described in \cite{hy}, we have for the pion distribution with the pion mass being neglected
 \begin{eqnarray}
 B(p_T) = {dN_{\pi}^B \over p_Tdp_Td\phi} = {C^2 \over 6} e^{-p_T/T} ,
 \label{4}
 \end{eqnarray}
which describes the observed low-\pt\ behavior of the bulk $(B)$ medium  having the parton distribution given by Eq.\ (\ref{3}).  In (\ref{4})  there is no $\phi$ dependence.
 
In addition to the thermal partons associated with the bulk medium described above, we now consider the enhanced thermal partons associated with semi-hard jets. A semi-hard scattering that can lead to a ridge must occur near the surface because otherwise the scattered parton would be absorbed in the interior of the medium and never emerge from the surface to generate any observable effect distinguishable from the isotropic background. The average direction of the scattered parton is normal to the initial surface, so let  us focus on such a representative semi-hard parton. Being massless it travels at $c$ and can reach the expanding surface before that surface deviates too much from the initial configuration. However, that is not when hadronization occurs. What happens to the semi-hard parton after it leaves the medium need not concern us here, since our present focus is on hadrons with $p_T\ ^<_\sim\ 2$ GeV/c. The emergent  semi-hard parton will generate shower partons that lead to hadrons at $p_T>2$ GeV/c. Of central importance to us is that in traversing the medium the semi-hard parton loses energy which leads to local enhancement of the thermal energy of the medium. Thermal equilibrium may or may not be established in less than 1 fm/c, but energy dissipated by the semi-hard parton that passes by has its effect in the  formation of ridge downstream. The medium must expand whether or not it can be described reliably by hydrodynamics starting at $\tau_0=0.6$ fm/c. By the time the density is low enough for hadronization to take place the geometry of the boundary of the medium may be very different from that of the initial configuration. Nevertheless, the region of the enhanced thermal partons is near the flow direction associated with the initial passage of the semi-hard parton, which is restricted by (1). We know from ridge phenomenology that the ridge particles are strongly correlated in $\phi$ with the trigger particles
 \cite{jp}. Although we have no trigger in the present problem, the semi-hard parton that generates a ridge specifies the direction to which the ridge particles are correlated.  Moreover, we expect a large number of semi-hard scattering to produce a continuum of ridges. Particles in that continuum are therefore in the range of $\phi$ corresponding to the angular window in which the semi-hard partons are emitted  with some round-off at the edges. The emission of semi-hard partons is not to be confused with the formation of soft hadrons in the ridges, which occurs much later; however, the ranges of $\phi$ of the two are strongly correlated.

There are many aspects in this non-perturbative problem that cannot be formulated rigorously. In the first place the virtuality of hard scattering is not narrowly restricted by any observable. Our main criteria are that it should be soft enough so that the yield of scattered partons is high, but hard enough so that the time of scattering is short.   The depth from the initial surface where the semi-hard scattering occurs can also vary over some range. The amount of energy loss depends on both the virtuality and depth, as well as on the density of the medium near the surface, which is not uniformly distributed in $\phi$. The non-uniformity is particularly severe near the tips of the almond-shaped overlap region. Thus the properties of the ridges formed at $\phi$ near $\Phi$ may be quite different from that near the reaction plane, which is where the most important contribution to the second harmonics in $\phi$ arises. The many complications that cannot be described rigorously force us to make simplifying assumptions that make possible our treatment without losing the essence of the problem. To that end, we make a sharp cutoff in $\phi$ at $\Phi$ for the ridge particles in the same spirit as in approximating the Woods-Saxon distribution by a sharp edge. We further use just one parameter $T'$ to summarize the uncalculable dynamical   process of ridge formation, and write 
\begin{eqnarray}
 q_0 {dN_q^{B+R} \over dq_Td\phi} = Cq_Te^{-q_T/T^{\prime}} ,  
 \quad |\phi|<\Phi, \quad |\pi-\phi|<\Phi .
 \label{5}  
\end{eqnarray}
This simple description  contains two main features: (a) it is a consequence of the semi-hard jets near the surface at early time, and thus has the $\phi$ dependence characterized by the pre-equilibrium partons; (b) the exponential $q_T$ dependence follows from the assumption of thermalization at later time, as in  (\ref{3}). The sharp edges in (\ref{5}) can be made smoother and the region in between may depart slightly from uniformity depending on density variation, but  the integrated effect of the anisotropy in $\phi$ should not depend sensitively on the approximations made, which have the virtue of rendering the final result in closed form.

 Both $T$ and $T'$ may be characterized by hydrodynamical flow and are properties of the pre-hadronization partons.  We do not treat  the problem between the initial and final states. We take the values of $T$ and $T'$ from the data to be discussed below.  Applying recombination to Eq.\ (\ref{5}) as before, we obtain 
 the  pion distribution for both  $B$ and   $R$ to be, for $|\phi|<\Phi$ and $|\pi-\phi|<\Phi$,
\begin{eqnarray}
B(p_T) + R(p_T, \phi) = {dN_{\pi}^{B+R} \over p_Tdp_Td\phi} = {C^2 \over 6} e^{-p_T/T^{\prime}} .
 \label{6}
\end{eqnarray}
The early-time consideration allows us only to determine the region of  $\phi$  that has the ridge without specifying the $q_T$ \dis\ of the partons at that time. The partons in the bulk and the ridge can take their time to thermalize without affecting in results in Eqs.\ (\ref{4}) and (\ref{6}). No rapid thermalization is required.
  
 From Eqs.\ (\ref{4}) and (\ref{6}) we obtain
 \begin{eqnarray}
 R(p_T, \phi)&=&R(p_T)  \Theta (\phi) ,  \nonumber
 \end{eqnarray}
 where
 \begin{eqnarray} 
 R(p_T)= {C^2 \over 6} e^{-p_T/T^{\prime}} \left(1 - e^{-p_T/T^{\prime\prime}} \right),  \label{7} \\
 \Theta (\phi) = \theta (\Phi - |\phi|) + \theta (\Phi - |\pi - \phi|) ,
 \label{8}  \\
 {1\over T^{\prime\prime} }={1\over T}-{1\over T'}= {\Delta T\over T T^{\prime} }, \qquad  \Delta T = T^{\prime} - T.
 \label{9}
 \end{eqnarray}
  Experimental study of the ridges in triggered events in central Au+Au collisions shows that $\Delta T$ is in the range of 40-50 
 MeV \cite{jp}.  
 We take $\Delta T=45$ MeV in the following.
 
 We note that the values of $C$ in Eqs.\ (\ref{3}) and (\ref{5}) are the same so that $R(p_T)\rightarrow 0$ as $p_T\rightarrow 0$. Since partons at $q_T=0$ do not flow, only the $q_T>0$ partons can lead to $\phi$ anisotropy, so $B+R$ cannot be distinguished from $B$ at $p_T=0$. There being no data on the ridge at $p_T\approx 0$, setting $C$ to be the same in Eqs.\  (\ref{3}) and (\ref{5}) is a reasonable assumption. 
 
\section{Elliptic Flow}
 
 The second harmonic in the $\phi$ distribution is \cite{jo}
 \begin{eqnarray}
 v_2(p_T) = \left< \cos 2 \phi \right> = {\int^{2\pi}_0 d\phi \cos 2 \phi\ dN/p_Tdp_Td\phi \over \int^{2\pi}_0 d\phi\ dN/p_Tdp_Td\phi}  . \label{10}
\end{eqnarray}
When we substitute for $dN/p_Tdp_Td\phi$ the general expression in terms of $B(p_T)$ and $R(p_T,\phi)$, we obtain 
 \begin{eqnarray}
 v_2(p_T, b)  &=& {\int d\phi \cos 2 \phi R(p_T, \phi) \over \int d\phi \left[B(p_T)+R (p_T, \phi) \right]}  \nonumber \\
  &=& {\sin 2 \Phi (b) \over \pi B(p_T)/R(p_T) + 2 \Phi (b)} . 
\label{11}
\end{eqnarray}
When Eq.\ (\ref{7}) is rewritten  as
\begin{eqnarray}
 R(p_T) =  {C^2 \over 6} e^{-p_T/T} \left(e^{p_T/T^{\prime\prime}} -1\right) ,
\label{13}
\end{eqnarray}
we have
\begin{eqnarray}
B(p_T)/R(p_T)=(e^{p_T/T''}-1)^{-1} ,  \label{14}
\end{eqnarray}
which  is large at small $p_T$, and dominates in the denominator of Eq.\ (\ref{11}). Thus we obtain for $p_T<0.5$ GeV/c
\begin{eqnarray}
 v^{\pi}_2(p_T, b) \simeq {p_T \over \pi T^{\prime\prime}} \sin 2 \Phi(b) . 
\label{15}
\end{eqnarray}
This is a simple formula that describes both the $p_T$ and centrality dependences at small $p_T$.

Note that $R(p_T)$ in Eq.\ (\ref{13}) is proportional to $B(p_T)$ in Eq.\ (\ref{4}), so   $C^2$ as well as the exponential factor are canceled out in Eq.\ (\ref{14}).  
Such factors can depend on centrality, but are irrelevant to $ v^{\pi}_2(p_T)$.  
$T''$ may have some centrality dependence, but it would be mild compared to that of the factor $\sin 2\Phi(b)$ that depends on $b$ strongly and explicitly.  The peak of $\sin 2\Phi(b)$ at $\Phi = \pi/4$ corresponds to $b = \sqrt{2}R_A$, which in turn corresponds to $\sim$50\% centrality for any colliding nuclei.  The data indeed shows a maximum of $v^{\pi}_2$ at that centrality for small \pt\ \cite{ja1} (see Fig.\ 1).  
From the $\pi^+$ \dis\ at 40-50\% \cite{ssa}, a fit of the exponential behavior in the range $0.5<p_T<2.7$ GeV/c gives $T=0.287$ GeV. 
Using that value of $T$ and 0.045 GeV for $\Delta T$, we obtain from Eq.\ (\ref{9})
\begin{eqnarray}
  T^{\prime\prime} = 2.12\, {\rm GeV} .
\label{16}
\end{eqnarray}
It then follows from Eq.\ (\ref{15}) that at small \pt\ the maximum value of $v_2^{\pi}$ (at 50\% centrality), divided by \pt, is
\begin{eqnarray}
  {v_2^{\pi}(p_T, b)\over p_T}={1\over \pi T''}=0.15\ ({\rm GeV/c})^{-1}
\label{17},
\end{eqnarray}
which agrees  with the data  that shows $v_2=0.075$ at $p_T=0.5$ GeV/c. This is the first quantitative demonstration that $v_2$ and ridge phenomenologies are related.
For higher $p_T$  the  approximation  given in Eq.\ (\ref{15}) is not good enough, so the full expression in Eq.\ (\ref{14}) should be used in (\ref{11}). The result is shown by   the thick line in Fig.\ 1 (left) for 50\% centrality, and agrees well with the data \cite{ja1} indicated by the blue stars for 40-50\% centrality. 

 \begin{figure}[tbph]
\centering
\includegraphics[width=6in]{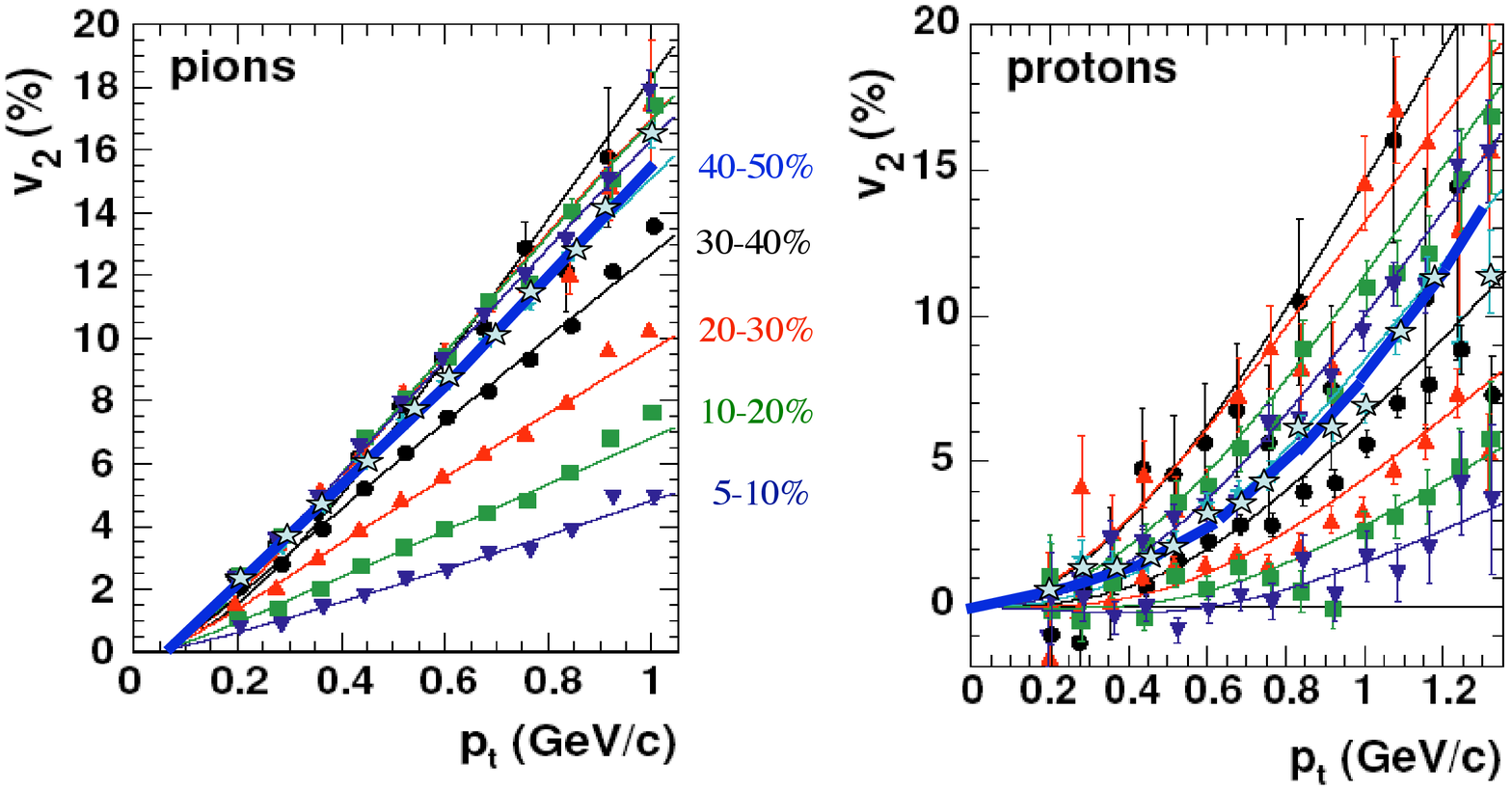}
\caption{Calculated $v_2^{\pi}(p_T)$ and $v_2^p(p_T)$ for 50\% centrality are indicated by the thick blue lines. The data are from \cite{ja1} for Au-Au collisions at 200 GeV, together with the light lines from that reference.}
\end{figure}

It should be remarked that data on the ridges are for $p_T>2.5$ GeV/c \cite{jp}; in that $p_T$ range the \dis\ can be described by either Eq.\ (\ref{6}) or (\ref{7}), giving essentially the same $T'$ with errors large enough to render $\Delta T=$40-50 MeV. Our use of $\Delta T=$45 MeV for smaller \pt\ in the absence of relevant data is an optimistic extrapolation that allows us to make an immediate connection between the ridge data and $v_2$.

For proton production the same formalism of recombination applies, as for pion, and the ridge consideration is the same. To take the mass effect into account we do what is usually done for hadronization, i.e.,  replacing all exponential forms, such as $e^{-p_T/T}$ for the bulk, by $\exp[-(m_T-m)/T]$, where $m_T=(p_T^2+m^2)^{1/2}$ and $m$ the proton mass. It then follows that at small \pt\ the  $v_2$ formula for proton is
\begin{eqnarray}
v_2^p(p_T,b)= {p_T^2\over 2\pi m T''}\sin 2 \Phi(b) .   \label{18}
\end{eqnarray}
With the same $T''$ as in Eq.\ (\ref{15}) we have  universal behavior of the slope of $v_2^h(p_T,b)$ versus the transverse kinetic energy, $E_T$, which is $p_T$ for pion and $p_T^2/2m$ for proton
\begin{eqnarray}
\left. {\partial v_2^h(p_T,b)\over \partial E_T}\right|_{p_T<0.5}= {1\over \pi T''}\sin 2 \Phi(b)    \label{19}
\end{eqnarray}
that is independent of the hadron type $h$. 
This universality is a consequence of $T''$ being a property of the partons in the ridges before hadronization. Thus 
 it trivially satisfies the  quark number scaling at low \pt, for which both $v_2^h$ and $E_T$ are scaled by the quark number $n_q$ \cite{mo,ad}.

For $0<p_T<1.2$ GeV/c we show by the thick blue line in Fig.\ 1 (right) the calculated $v_2^p(p_T)$ for $\Phi=\pi/4$ corresponding to 50\% centrality; it agrees well with the data for 40-50\% \cite{ja1}.
For more peripheral collisions   Eq.\ (\ref{18}) is inadequate to describe proton production   because less thermal partons render the joint $uud$-distribution not factorizable and the effective value of $T$  in the bulk is reduced. That leads to a decrease of $B(p_T)/R(p_T)$ and the second term  in the denominator of Eq.\ (\ref{11}) cannot be ignored. The consequence is that $v_2^p(p_T)$ continues to increase for larger $b$, a feature that will be investigated in  detail  in the future. For  the main range $0<b<\sqrt 2R_A$ our simple treatment described above is adequate.

If we define the scaled impact parameter to be $\beta=b/2R_A$, the explicit factor that describes the centrality dependence is 
\begin{eqnarray}
f(\beta)=\sin 2(\cos^{-1} \beta),  \label{20}
\end{eqnarray}
which is plotted  in Fig.\ 2. It agrees with the Au-Au data on $v_2$  \cite{ca},  normalized to unity at maximum. Since $f(\beta)$ does not depend  on $R_A$ explicitly, one expects it to agree with the Cu-Cu data also. Evidence is given  in Ref.\ \cite{rh}.

 \begin{figure}[tbph]
\centering
\includegraphics[width=4in]{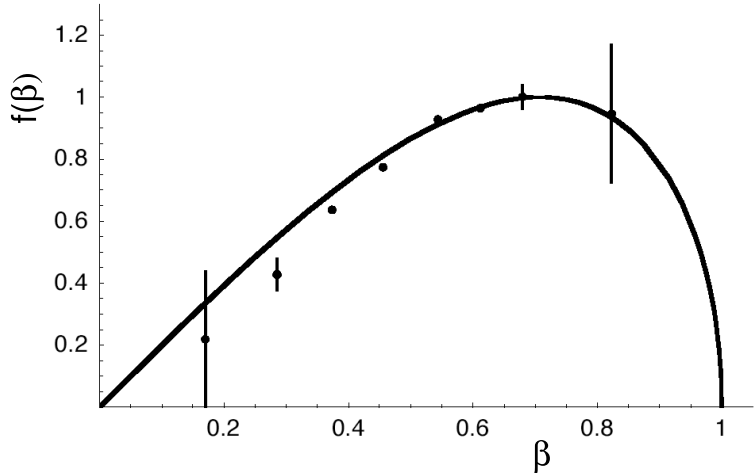}
\caption{Centrality dependence of $v_2$ normalized to 1 at maximum. Data are $v_2$ for Au-Au collisions at 130 GeV \cite{ca}, scaled to unity at b=10fm.}
\end{figure}

The next question concerns the dependence of $T''$ on energy and density. Properties of the ridge have so far been studied with the help of triggers, and our knowledge about ridges with unspecified semi-hard parton momenta is rudimentary. Our assumption that $T''$ is a constant is a reasonable first approximation; however, we expect it to be improved upon further investigation.

The above consideration applies only at midrapidity. In forward region the semi-hard scattering involves partons at larger $x$, but not too large because of recombination \cite{hy2}. Lower parton density means that there are less phantom jets, so   the ridge effect is reduced with increasing $\eta$, resulting in diminishing $v_2$.

\section{Summary}

We have shown that the observed azimuthal anisotropy at low \pt\  can be reproduced by a consideration of the  ridges whose formation is due to semi-hard scattering of partons. 
The hadrons in the ridges are produced by recombination of enhanced thermal partons near the surface of the medium. The centrality dependence is described mainly by a geometrical factor, $\sin 2\Phi(b)$. The initial slope in the $E_T$ dependence of $v_2$ is essentially $1/\pi T''$ for both pion and proton, where $T''$ is related to parameters in ridge phenomenology. This is the first time that a connection is made between properties of elliptic flow and of ridges due to semi-hard scattering.
No fast thermalization is required, and no hydrodynamical flow from early times is assumed. We do not exclude the relevance of hydrodynamics at some later stage of the expansion, but it is not needed to derive the main features of $v_2$ from ridge data. 

This work was supported, in part, by the U.\ S.\ Department of Energy under Grant No. DE-FG02-96ER40972.

\end{document}